\documentclass[twocolumn]{svjour3}

\usepackage[T1]{fontenc}
\usepackage{mathptmx}
\usepackage[utf8]{inputenc}
\usepackage{graphicx}
\usepackage{url}
\usepackage{hyperref}
\usepackage{amssymb}

\newcommand{\mytitle}{Seamless Paxos Coordinators}

\hypersetup{
     pdftitle={\mytitle}, 
     pdfauthor={Gustavo M. D. Vieira, Islene C. Garcia, Luiz E. Buzato},
     pdfdisplaydoctitle=true,
     hidelinks
}

\journalname{Cluster Computing}

\title{\mytitle}

\author{Gustavo M. D. Vieira \and Islene C. Garcia \and
        Luiz E. Buzato}

\institute{G.  M.  D.   Vieira  \at
           DComp, CCTS, Federal University of São Carlos, Brazil\\
           \email{gdvieira@ufscar.br}
           \and I.  C.  Garcia {and} L.  E.  Buzato \at
           Institute of Computing, University of Campinas, Brazil
           \and I.  C.  Garcia \at
           \email{islene@ic.unicamp.br}
           \and L.  E.  Buzato \at
           \email{buzato@ic.unicamp.br}}

\date{}

\begin{document}

\maketitle

\begin{abstract}
  The Paxos algorithm requires a single correct coordinator process to
  operate.  After  a failure, the  replacement of the  coordinator may
  lead to  a temporary  unavailability of the  application implemented
  atop  Paxos.  So  far,  this unavailability  has  been addressed  by
  reducing the coordinator replacement  rate through the use of stable
  coordinator selection algorithms.  We have observed that the cost of
  recovery of the newly elected  coordinator's state is at the core of
  this  unavailability  problem.   In  this  paper we  present  a  new
  technique to manage coordinator replacement that allows the recovery
  to  occur  concurrently  with  new consensus  rounds.   Experimental
  results  show  that our  seamless  approach  effectively solves  the
  temporary unavailability problem, its adoption entails uninterrupted
  execution of  the application. Our solution  removes the restriction
  that the  occurrence of coordinator replacements is  something to be
  avoided, allowing  the decoupling of the  application execution from
  the accuracy  of the mechanism  used to choose a  coordinator.  This
  result  increases the  performance of  the application  even  in the
  presence of failures, it is  of special importance to the autonomous
  operation of  replicated applications that have to  adapt to varying
  network conditions and partial failures.

\end{abstract}

\section{Introduction}

Total  order broadcast  primitives are  a critical  component  for the
construction   of  fault-tolerant   applications  based   upon  active
replication, aka state machine replication~\cite{lamport78,schneider90}.
The primitive guarantees that messages  sent to a set of processes are
delivered, in their turn, by all  the processes of the set in the same
total order.  A possible way of implementing total order broadcasts is
through  multiple  executions of  a  consensus  algorithm.  Thus,  the
performance of the total order  broadcast is directly dependent on the
performance  of the consensus  algorithm.  This  paper focuses  on the
performance  and adaptability  of Paxos~\cite{lamport98},  a consensus
algorithm that has been used  to support the construction of real-life
fault-tolerant    systems    such   as    Boxwood~\cite{maccormick04},
Chubby~\cite{burrows06}, and Spinnaker~\cite{rao11}.

Paxos  has  been designed  for  asynchronous  distributed systems,  it
relies on  a procedure executed  by a  key agent, the  coordinator, to
ensure its safety.  The algorithm  also guarantees liveness as long as
there is one, and only one, coordinator.  When Paxos is used to decide
multiple  instances  of consensus,  as  in  the  case of  total  order
broadcast,  the  coordinator  also   ensures  the  algorithm  performs
optimally,  reaching consensus  in  three communication  steps in  the
absence   of   failures~\cite{lamport98}.    Thus,   the   coordinator
effectively acts as a sequencer and processes all application messages
that need  to be  ordered; it  does so  by initiating  many concurrent
consensus instances and keeping track of their outcome.

As  any other process  of the  system, the  coordinator is  subject to
failures  that  eventually will  cause  its replacement.   Coordinator
\emph{replacement} is carried  out in two steps: a  new coordinator is
elected,  and  then  it  is  validated~\cite{lamport98}.   Coordinator
\emph{election} is handled by any unreliable leader election mechanism
that is equivalent  to an $\Omega$ failure detector~\cite{chandra96c}.
The  unreliability   of  the  election  means  that   it  allows  many
coordinators to  be changed  many times, but  it will select  a single
coordinator eventually.   Coordinator \emph{validation} is  crucial to
the  Paxos ability to  reach consensus  in three  communication steps,
ensuring that the new coordinator is  up to date with the state of all
active consensus instances.  To  achieve this, validation requires the
new  coordinator to  have its  role ratified  by a  majority  of Paxos
agents.   During  the ratification  process,  the  coordinator has  to
receive and process a potentially large prefix of the current state of
each member  of the  majority.  A newly  elected coordinator  can only
resume its  activities after the completion of  validation.  Thus, the
replacement  of a  coordinator  triggers a  costly  operation that  is
certainly   going   to   lead   to   a  temporary   halting   of   the
application~\cite{lamport98}.   Coordinator replacements are  bound to
happen  reasonably  often in  the  presence  of  partial failures  and
incomplete  or  inaccurate failure  detection.   So  the  root of  the
\emph{temporary unavailability problem} is  the fact that normal Paxos
operation  can only be  resumed after  a successful  validation.  This
problem is a  real concern for fault-tolerant systems  based on Paxos,
because its  performance becomes  dependent on the  error rate  of the
failure detector used in  the system; the characteristics of solutions
based  on   failure  detectors  are  analyzed  in   the  related  work
(Section~\ref{sec:related}).

In  this  paper, we  show  a  novel  solution to  the  \emph{temporary
  unavailability  problem}   that  stems  from   breaking  coordinator
validation  in  two concurrent  activities:  activation and  recovery.
Coordinator \emph{activation}  corresponds to the  actual ratification
of a coordinator by a majority.  We show that it is possible to reduce
the information necessary to activate  the new coordinator to a single
integer.  We show that the coordinator doesn't need to rebuild at once
its  complete  state,  from  information  gathered of  a  majority  of
processes, before it  can resume its work. In  fact, all a coordinator
needs  is to  discover  the highest  non-initiated consensus  instance
using the  local knowledge  of a majority  of processes.  This  can be
done using only a single exchange of fixed size messages, allowing the
new coordinator to resume operation in a very short time.

Coordinator  \emph{recovery} becomes  a secondary  task that  can take
much  longer to  finish.  The  coordinator  won't be  able to  deliver
messages  locally while  it  recovers,  but that  does  not block  the
progress  of  the  other  processes  of  the  application  during  the
validation. This  happens because the  gap present in  the coordinator
state is  not necessarily reflected  in the state of  other processes,
thus  they can  continue to  deliver requests  to their  clients.  The
result is a much briefer  coordinator validation whose time is limited
primarily by  the activation time.  The  coordinator's state recovery,
the longer step, occurs while  the coordinator is already managing new
consensus instances.   Moreover, the coordinator can  limit the impact
its  recovery has  on the  overall performance  of the  application by
limiting  the  amount of  communication  or  computation it  performs,
effectively  slowing down  or speeding  up its  recovery  as required.
From the point of view of the application our new procedure guarantees
coordinator    replacements   with    less    disruptive   performance
oscillations, namely,  \emph{seamless coordinator validations}.  These
validations  happen in  constant time,  making Paxos  performance less
susceptible to the unreliability of the failure detection mechanism.

The  reduction  of the  cost  associated  with  the replacement  of  a
coordinator  has  other  implications  for  the  research  on  failure
detectors for  Paxos.  Specifically,  seamless validation  removes the
restriction that  the occurrence  of coordinator replacements  must be
avoided.   Thus,  instead  of  deploying and  tuning  complex  failure
detectors,  it  is possible  to  use  a  very simple  leader  election
mechanism  to choose  a new  coordinator.  A  fairly imprecise  leader
election  procedure, but  one that  responds  fast to  failures or  is
simpler to implement, can be used without hindering the performance of
Paxos.   More important,  by decoupling  the performance  of consensus
from the  accuracy of the failure  detector, it is possible  to ensure
autonomous operation of replicated applications under variable network
conditions and failure patterns.

Experimental results confirm that  our concurrent validation procedure
guarantees  progress  with sustained  throughput  in  the presence  of
coordinator replacements caused by  both process and network failures.
While  the coordinator  replacements  happened, we  have observed  the
continuous operation of  the application, a clear  indication that our
validation procedure solves the temporary unavailability problem.  The
net result is that the proposed coordinator validation mechanism makes
coordinator replacement seamless to the application.

The   rest   of   the   paper    is   structured   as   follows.    In
Section~\ref{sec:paxos} we give an overview of the Paxos algorithm and
introduce the terms used  throughout the paper.  Section~\ref{sec:ocv}
discusses the  original coordinator validation procedure  of Paxos and
its       link      with       the      temporary       unavailability
problem. Section~\ref{sec:seamless} describes the seamless coordinator
validation     procedure      and     proves      its     correctness.
Section~\ref{sec:experiments} discusses the results of the experiments
carried  out to  compare  the original  with  the seamless  validation
procedure.    Section~\ref{sec:related}    describes   related   work.
Section~\ref{sec:conclusion} provides concluding remarks.

\section{Paxos}
\label{sec:paxos}

Informally, the \emph{consensus} problem consists in each process of a
distributed  system  proposing  an  initial value  and  all  processes
eventually  reaching  a unanimous  decision  on  one  of the  proposed
values.   The Paxos  algorithm is both  a solution  to the  consensus
problem and a  mechanism for the delivery of  totally ordered messages
that       can       be       used       to       support       active
replication~\cite{lamport78,schneider90}.  In  this section we  give a
summarized  description  of  Paxos  and  make explicit  the  key  role
performed by the coordinator.   Full descriptions of the algorithm can
be found in~\cite{lamport98,lamport06a}.

\subsection{Core Algorithm}

Paxos is specified  in terms of roles and agents;  an agent performs a
role. Different implementations of Paxos may choose different mappings
between  agents and  the actual  processes that  execute  them. Agents
communicate exclusively via message exchanges.  The usual asynchronous
crash-recovery computation model is assumed. The roles agents can play
are:  a \emph{proposer}  that can  propose values,  an \emph{acceptor}
that chooses a single value, or a \emph{learner} that learns what value
has been  chosen.  To solve  consensus, Paxos agents  execute multiple
\emph{rounds},  each round  has a  \emph{coordinator} and  is uniquely
identified by a positive integer.  Proposers send their proposed value
to the  coordinator that tries  to reach consensus  on it in a round.
The coordinator is responsible for  that round, and is able to decide,
by applying a local rule, if  any other rounds were successful or not.
The local rule of the coordinator is based on quorums of acceptors and
requires that at  least $\lfloor N/2 \rfloor +  1$ acceptors take part
in  a  round, where  $N$  is  the total  number  of  acceptors in  the
system~\cite{lamport06a}.   Each round  progresses through  two phases
with two steps each:
\begin{itemize}
\item In  Phase 1a  the coordinator sends  a message  requesting every
  acceptor  to participate  in  round $r$.   An  acceptor accepts  the
  invitation if it has not already accepted to participate in round $s
  \geq r$, otherwise it declines the invitation by simply ignoring it.
\item In  Phase 1b,  every acceptor that  has accepted  the invitation
  answers  to the  coordinator with  a reply  that contains  the round
  number and  the value of the last  vote it has cast  for a
  proposed value, or \textsl{null} if it has never voted.
\item  In Phase  2a,  if the  coordinator  of round  $r$ has  received
  answers from  a quorum of acceptors,  it analyzes the  set of values
  received  and picks  the single  value  $v$ with  the highest  round
  number.  It then asks the acceptors  to cast a vote for $v$ in round
  $r$, if $v$ is not  \textsl{null}, otherwise the coordinator is free
  to pick any value and picks the value proposed by the proposer.
\item In Phase  2b, after receiving a request  from the coordinator to
  cast a vote, acceptors can either  cast a vote for $v$ in round $r$,
  if they  have not  voted in  any round $s  \geq r$,  otherwise, they
  ignore the vote  request.  Votes are cast by  sending them and their
  respective round identifiers to the learners.
\item Finally, a  learner learns that a value $v$  has been chosen if,
  for some round  $r$, it receives Phase 2b messages  from a quorum of
  acceptors announcing that they have all voted for $v$ in round $r$.
\end{itemize}

\begin{figure}[htbp]
  \begin{center}
    \includegraphics[width=6cm]{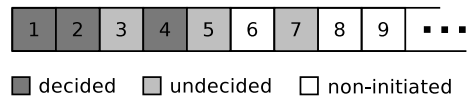}
  \end{center}
  \caption{Local View of an Agent}
  \label{fig:local}
\end{figure}

This description of the algorithm  considers only a single instance of
consensus.  However,  Paxos also  defines a  way to  deliver a  set of
totally ordered messages.  The order of delivery of  these messages is
determined by a sequence of  positive integers, such that each integer
maps to a consensus instance.   Each instance $i$ eventually decides a
value $v$ and  this value is the message (or  ordered set of messages)
to be delivered  as the $i$th message of the  sequence.  The value $v$
is input by  the proposers, and they can either  select a suitable $i$
from their local view of the  instance sequence or ask the coordinator
to select $i$  from its view.  Each consensus  instance is independent
from the  others and  many instances  can be in  progress at  the same
time.   In fact,  for any  agent  its local  view  of the  set of  all
instances can be  divided in three proper  subsets: the \emph{decided}
instances, the  \emph{undecided} instances that were  initiated (Phase
1a)   and  the   infinite  set   of  \emph{non-initiated}   instances.
Figure~\ref{fig:local} shows an example of the status of the consensus
instances as  seen by an  agent.  In this  example the set  of decided
instances is $\{1, 2, 4\}$, the set of undecided instances is $\{3, 5,
7\}$ and the  set of non-initiated instances  is $\mathbb{N} \setminus
\{1, 2, 3, 4, 5, 7\}$.

\subsection{Stable Storage Requirements}

Paxos assumes  a process  failure model where  agents crash  and later
recover.  When a process crashes, it  loses all state it has stored in
its  local volatile  memory.  Unfortunately,  key information  must be
restored  exactly  as  it  was  before  the  crash  to  guarantee  the
correctness  of the  algorithm. Thus,  parts  of the  local state  are
recorded  into  stable  storage~\cite{lampson81}.   Access  to  stable
storage  is  usually  slow,  so   its  use  must  be  minimized.   The
coordinator must store the value of the last round it has started, say
$crnd_c$,    to   ensure    it    won't   start    the   same    round
twice~\cite{lamport06a}.  Similarly, each acceptor must save in stable
storage:

\begin{itemize} 
\item $rnd_a$: the last round they have taken part (Phase 1a);
\item $vrnd_a$: the last round where they have  cast a vote;
\item $vval_a$: the value of the vote cast in $vrnd_a$  (Phase 2a).

\end{itemize}

The stable storage requirements for the set of  consensus instances in
Paxos are the same for a single instance, but multiplied by the number
of  instances.  Thus,  each agent  must store  an array  of instances,
where  for  each  instance  $i$ it  records  $rnd_a[i]$,  $vrnd_a[i]$,
$vval_a[i]$  and  $crnd_c[i]$.  Additionally,  the  learner agent  may
store $dval_l[i]$, the  value decided in instance $i$,  but this isn't
strictly  necessary as  a new  successful  round will  yield the  same
value.  Usually, all agents are implemented in each process and agents
may  use   the  information  stored  by  other   agents  to  implement
optimizations.  For instance, a  coordinator can inform proposers that
their selected  instance number $i$ is already  decided, or similarly,
acceptors can inform a coordinator that an instance $i$ it is about to
start is already decided.

\subsection{Liveness and Safety}

In  Paxos, any  process  can act  as  the coordinator  as  long as  it
correctly chooses  a value,  if any, that  has been proposed  in Phase
2a. There can be only one active coordinator at any given time for the
algorithm  to  ensure  progress.   If  two  or  more  processes  start
coordinator  agents,  the  algorithm  can  stall  while  the  multiple
coordinator candidates  cancel each other rounds  with fast increasing
round numbers.  For this reason,  the liveness of the algorithm relies
on  an   unreliable  $\Omega$  failure  detector.    Safety  is  never
compromised, even if multiple coordinators, including none, are active
at any  time.  However,  the $\Omega$ implemented  needs to  be robust
enough to guarantee  that only a single coordinator  is active most of
the time.

\section{Original Coordinator Validation}
\label{sec:ocv}

If one considers the sequence of consensus instances necessary for the
delivery of  totally ordered  messages, it is  possible to  reduce the
five communication  steps required  by Phases  1 and  2 to  only three
communications  steps,   by  running  Phase   1  only  once   for  all
non-initiated  instances.   We  call   this  factorization  of  phases
\emph{validation} and it is carried out immediately after the election
of a  coordinator.  In  this section  we describe  in more  detail how
validation     is     performed      in     the     original     Paxos
specification~\cite{lamport98}.

During validation a coordinator selects  a round number $r$ and starts
\emph{all} consensus instances at the same time with a single message,
as the  Phase 1a  message carries  only the round  number.  If  $r$ is
large enough,  acceptors will  respond to this  message with  a finite
number  of Phase 1b  messages with  the actual  votes and  an infinite
number of  Phase 1b messages with  no votes.  Lamport~\cite{lamport98}
notes that only  the finite set of messages  containing an actual vote
needs to be sent back to  the coordinator, framed in a single physical
message.  No message has to be sent to the coordinator for each of the
infinite instances  that have had  no vote yet.  When  the coordinator
receives  this combined  message  for  each process  in  a quorum,  it
processes all  Phase 1b messages  received and it  \emph{assumes} that
the infinitely  many omitted messages correspond to  Phase 1b messages
with  no  vote.   All  messages  received  or  presumed  voteless  are
processed as usual  and a suitable value will be  selected to be voted
for each  instance, or the instance  will be marked  free (no previous
value) and  will be used when  necessary.  This way  a coordinator can
start  the Phase  2 of  any free  instance as  soon as  it  receives a
proposal,  and consensus  for this  instance can  be reached  in three
communication steps~\cite{lamport06a}.

This validation procedure requires the coordinator to learn the status
of  all decided  and  undecided  consensus instances  of  a quorum  of
acceptors  to   determine  the   exact  identities  of   the  infinite
non-initiated consensus instances. So,  the combined state of a quorum
of  acceptors represents the  state footprint  a new  coordinator must
recover  to be  able to  start  passing new  consensus instances.   To
reduce  the  footprint  of  the  recovery state,  it  is  possible  to
determine  a point  $d_c$ in  the instance  sequence, as  seen  by the
coordinator, such that all instances $i$, with $i \leq d_c$, belong to
the decided  set.  The point  $d_c$ doesn't necessarily  determine all
instances in the decided set,  it only captures the local knowledge of
the coordinator.  The coordinator can then send $d_c$ to the acceptors
to  inform  what it  knows  about the  decided  set.   This way,  each
acceptor needs  only to  send back to  the coordinator  information on
consensus      instances     with     identifiers      larger     than
$d_c$~\cite{lamport98}.  This procedure reduces  the size of the state
the coordinator has to receive and process, but it can still represent
a very large state that must be fully recovered so the coordinator can
(1)  discover all consensus  instances that  have received  votes from
acceptors and (2)  use this information to infer  the set of instances
that \emph{have not} received  votes from acceptors.  Moreover, before
the coordinator can  complete Phase 1 for all  consensus instances, it
must have received answers from at least a quorum of acceptors.  While
this  happens, the  coordinator  remains blocked  and  no progress  is
possible; the whole application becomes unavailable.

\section{Seamless Coordinator  Validation}
\label{sec:seamless}

Our proposal  for a  seamless coordinator validation  is based  on the
observation  that   validation  can   be  broken  in   two  concurrent
activities:  activation  and recovery.   Activation  is the  procedure
where  acceptors  inform  the  newly  elected  coordinator  about  the
instances they  have not voted.   Recovery is the procedure  where the
coordinator's view  of the consensus  instances is updated,  it learns
the  outcome  of  decided  instances  and  initiates  rounds  for  the
undecided  ones.  This compound  view of  the validation  procedure is
interesting because only activation  is required to be finished before
a coordinator  can resume its activities.   Recovery, although necessary,
does   not  pose  any   restriction  on   the  coordinator's   use  of
non-initiated consensus instances.  This happens because a coordinator
doesn't need to immediately  start the consensus instances that belong
to the  undecided set.  For  these instances, the  coordinator doesn't
know whether it can instantly input  a value or not, as a consequence,
it can  learn their  status later, during  recovery.  In order  to use
this fact  to create  a more efficient  validation we have  devised an
activation  procedure that  avoids  the transfer  of  the finite,  but
possibly very large, set  of decided and undecided consensus instances
that  make up  the recovery  state.  Before  we can  describe  how the
coordinator can achieve this economy in the state transferred from the
acceptors,  it is  useful to  understand  the views  of the  consensus
instances held by the coordinator and the acceptors.

A coordinator must  be able to produce an  ever increasing sequence of
round numbers  to be used  in consensus instances.  These  numbers are
distinct and increasing  but they need not to  be sequential. In fact,
they are completely partitioned among the processes in an even way; so
they  are never  sequential.  Thus,  for instance  $i$,  a coordinator
picks a round  number to be any round  number larger than $crnd_c[i]$,
but not necessarily $crnd_c[i] +  1$.  From this simple observation it
is easy to see that if  the coordinator only records the largest round
number initiated for \emph{all} instances, it is guaranteed to be able
to always  choose a  larger round number  for any  individual instance
when necessary.   In this  case, the stable  storage footprint  of the
coordinator can  be reduced to the  space necessary to  store a single
integer $crnd_c$, no matter how  many instances of consensus were ever
initiated by  it.  Clearly, the  coordinator still must keep  track of
the progress of  the rounds it initiates, including  the round numbers
of  the rounds  in progress,  but this  information may  be  stored in
volatile memory.   This simple observation  makes clear the  fact that
the coordinator  doesn't concern itself  with the proposed  or decided
values of consensus, but only  with the proper initiation and progress
of rounds.

An acceptor, however,  needs to keep a persistent  history of the last
round  it  took  part,  for  every instance.   Each  instance  $i$  is
initially  inactive  and  belongs  to  the  non-initiated  set,  their
corresponding variables ($rnd_a[i]$, $vrnd_a[i]$ and $vval_a[i]$) have
been initialized as \emph{null}.  As Paxos progresses, values computed
by the agents are stored in  the fields of the consensus instances and
they pass to the set  of active but undecided instances. Eventually, a
consensus is reached for an instance and it is promoted to the decided
set.  As  the number of  instance identifiers picked by  the proposers
from  the  set of  positive  integers is  finite,  it  is possible  to
establish  a  point $f_a$  in  the  instance  sequence, as  viewed  by
acceptor  $a$, such that  every instance  $i$, $i  \geq f_a$,  has not
received a  vote yet.  It is  possible to find  some instances smaller
than $f_a$  that have also  not received a  vote yet, but we  know for
sure  that all  instances larger  than or  equal to  $f_a$  have never
received  a  vote.   For  example,  in the  local  state  depicted  in
Figure~\ref{fig:local} we have $f_a = 8$.

\subsection{Activation Procedure}
\label{sec:sa}

The  seamless  coordinator  validation   is  based  on  an  activation
procedure that determines a point  $f_Q$ of the global Paxos consensus
history using points $f_a$ of the local histories of each acceptor $a$
of  a  quorum $Q$.  The  detailed  steps  executed by  the  activation
procedure are as follows:

\begin{figure}[htbp]
  \begin{center}
    \includegraphics[width=6.1cm]{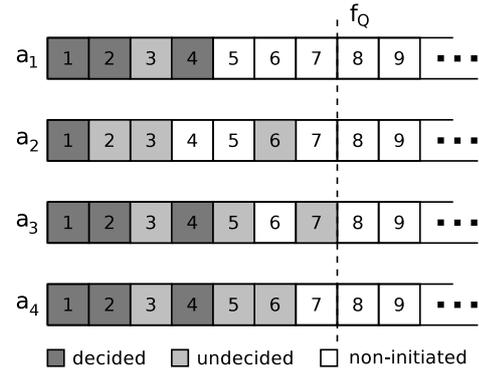}
  \end{center}
  \caption{Global View as Observed by a Coordinator}
  \label{fig:global}
\end{figure}

\begin{enumerate}
\item The coordinator sends an Activation Phase 1a message,
  with round number $r$ starting \emph{all instances}.
\item  \label{sa2}  When an  acceptor  $a$  receives  this message  it
  computes its
  $f_a$.  If $r$ is larger than the last round number used in another
  \emph{activation}  or there  was  no previous  activation, then  $a$
  sends  a single Activation  Phase 1b  message containing  its $f_a$,
  meaning that  it is sending Phase  1b messages for  all instances $i
  \geq f_a$ and \emph{only for these instances}.
\item As soon as the coordinator  has received Activation
  Phase 1b messages from a  quorum $Q$ of acceptors, it computes $f_Q$
  to be  the largest of  the $f_a$ received,  for each $a \in  Q$.  It
  then considers that it has received a Phase 1b message with no votes
  from all acceptors in $Q$ for  instances $i \geq f_Q$, and from this
  point on it proceeds as the original Paxos.
\end{enumerate}

Figure~\ref{fig:global} shows an example of the activation process for
four acceptors $a_1$, $a_2$, $a_3$ and $a_4$. Assuming all of them are
able to take  part in the activation, they  compute $f_a$ respectively
as $f_{a_1} = 5$, $f_{a_2} = 7$,  $f_{a_3} = 8$ and $f_{a_4} = 7$. The
coordinator computes $f_Q = 8$  and ends its activation.  The instance
$f_Q$ is the first consensus instance the coordinator can later expect
a quorum to respond to its Phase 2a message and to decide consensus in
only three communications steps.

The   seamless   coordinator   activation  presented   here   requires
considerably less information  to be propagated from  the acceptors to
the  coordinator, despite  the preservation  of the  communication and
time complexity of the original Paxos~\cite{deprisco00}.  It takes one
broadcast from  the coordinator  containing the  round number  and $Q$
unicasts from  the acceptors  to the  coordinator containing  a single
integer  $f_a$.    This  contrasts   with  the   original  coordinator
validation~\cite{lamport98}  where  the  activation and  recovery  are
handled  sequentially.  In  the  original  validation the  coordinator
broadcast is  answered by $Q$  unicasts containing all  previous votes
for  consensus   instances  $d_c   <  i  <   f_a$,  as   described  in
Section~\ref{sec:ocv}.  Each vote contains,  besides the round number,
the contents  of the  actual application messages  (or ordered  set of
messages)  voted  in  one  specific consensus  instance.   It  is  not
difficult to see that the  handling of the transmission, reception and
processing  of  these  much  larger messages  can  have  a
considerable  cost for  Paxos.  More  important, while  the sequential
validation is  underway Paxos  stops delivering  application messages,
causing the temporary unavailability problem.

\subsection{Correctness}

The correctness of the seamless coordinator activation is derived from
the correctness of individual  Paxos consensus instances.  Although in
its first step the coordinator  initiates many instances at once, each
one of them complies strictly  with Paxos protocol and with the proofs
contained  in~\cite{lamport98}.  So,  in this  section, we  sketch the
proof that the  activation procedure we have devised  does not perform
any operation forbidden by the original Paxos.

Steps 1 and  2 of the activation  procedure (Section~\ref{sec:sa}) are
functionally  identical to  the original  Paxos algorithm,
the difference is only in the  content of the messages exchanged.  The
change introduced to the messages just makes explicit that an acceptor
must keep track  of all Activation Phase 1b messages  it has responded
to,  and must  refrain from  taking part  in activations  with smaller
round  numbers.    This  is  consistent  with   the  observation  that
activation  is  just  the  execution  of Phase  1  for  all  consensus
instances.     As    the    $f_a$   point    is    uniquely    defined
(Section~\ref{sec:seamless}),  each acceptor  will respond  Activation
Phase  1a  messages  with  a sufficiently  large  round  number.   The
coordinator is able to eventually receive a non empty set of responses
from a  quorum, if it  is unique  and keeps starting  activations with
increasing round numbers.

In Step 3, we must show that the determination of $f_Q$ allows for the
correct   determination  of   the  set   on   non-initiated  consensus
instances. In any Paxos round, the  coordinator is only free to set an
arbitrary value to  an instance if it receives  only \emph{null} votes
from  all acceptors  in a  quorum.  For  any given  acceptor  $a$, the
coordinator considers  that it has  received a \textsl{null}  vote for
all instances $i  \geq f_a$.  The coordinator receives  answers from a
quorum $Q$  and establishes the point  $f_Q$ to be  the largest $f_a$,
for  all acceptors  $a  \in  Q$.  It  is  easy to  see  that only  for
instances at  least as large as  $f_Q$ a full  quorum of \textsl{null}
votes is received. All instances smaller than $f_Q$ will miss at least
one vote  to complete  a quorum.  The  coordinator then can  treat all
instances $i \geq  f_Q$ as started and free to  use.  This leaves many
instances  $i < f_Q$,  that are  not yet  decided, from  the acceptors
where $f_a < f_Q$. These  instances will be treated normally later, as
they are not required for the coordinator operation.

\section{Experimental Evaluation}
\label{sec:experiments}

The seamless coordinator validation  allows activation and recovery to
occur  concurrently.   It is  reasonable  to  suppose  that the  added
concurrency  will reduce  the  time a  coordinator  is blocked  during
validation, allowing Paxos to work without interruption. Moreover, the
time required  by seamless validation  is sufficiently short  that the
error  rate of  the failure  detector  isn't crucial  for the  overall
performance of the system. To  assess this hypothesis we have designed
two sets  of experiments.  The first set  investigates the performance
of  the original and  seamless validations  during executions  where a
very simple non-stable failure detector is used. In this experiment we
induce  the  failure and  recovery  of  a  specific process  that  the
non-stable failure  detector will  select as coordinator.   The second
set  of  experiments  compares   the  performance  of  the  validation
procedures  in the  more realistic  situation where  a  stable failure
detector is  used. Here, we induce  a partition of  the network around
the current coordinator,  isolating it from the rest  of the processes
for a  certain period of time.   As soon as the  partition is removed,
the \emph{stable} failure detector  will select as the coordinator the
same process that was the coordinator before the partition occurred.

The  results  of both  sets  of  experiments  show that  the  seamless
coordinator  validation  guarantees  not  only  that  Paxos  does  not
interrupt its delivery of ordered messages during the replacement of a
coordinator but also  that it makes the application free  from the ill
effects  caused by  inaccurate failure  detection,  as can  be seen  in
Figures~\ref{fig:crash} and  \ref{fig:net}.  In the remainder  of this
section we  further detail  the experiments, with  an emphasis  on the
components and parameters they have in common.

\subsection{Method}

Our tests were made using Treplica, a modular replication toolkit that
implements Paxos  and Fast  Paxos~\cite{buzato09,vieira08a}.  Treplica
has  been  designed   to  be  easily  instrumented   to  generate  the
performance  indicators  necessary  to   assess  Paxos.   The  toolkit
provides  a  programming interface  that  allows  the construction  of
applications that  adhere to  the state machine  replication approach.
In  this  model an  application  is  the collection  of  deterministic
\emph{replicas}   that   change   their  state   by   processing   the
totally-ordered messages delivered by Treplica.

To assess the seamless validation the experiments compare the relative
performance of  the two  coordinator validation procedures.   Thus, to
minimize any possible effect of  execution of the application upon the
performance measurements we have  implemented a very simple replicated
application: a  persistent table with integer  entries.  Table entries
can be  read and written.  The  workloads used in all  the experiments
are   exclusively   composed    of   \emph{write   operations}.    The
\emph{workload} is generated by selecting the number of operations per
second (op/s) the application receives  from its clients, the workload
generators,  while maintaining  the size  and execution  time of  each
individual  operation   constant.   The  performance  metric   of  the
experiments is the application \emph{throughput}, it measures the rate
of  operations   per  second   (op/s)  effectively  executed   by  the
application.  Replicas  and workload generators share  the same hosts,
but care  has been  taken to  ensure that  the load  generation wasn't
competing with  the replicas processing  capacity and that  the chosen
workload   was  kept   constant   throughout  the   duration  of   the
experiment. Indeed,  we guaranteed  that the workload  generators were
capable of creating 1000 op/s without disturbing the processing of the
replicated system---in  fact, we established that  for workloads below
10,000 op/s the  workload generator consumed less than 1\%  of CPU and
had a memory footprint smaller than 10 MB.

The  experiments  were  carried  out   in  a  cluster  with  16  nodes
interconnected by  a 1Gbps  Ethernet switch.  Each  node has  a single
Intel Xeon E5620  processor (2.4 GHz, 8 threads), 12 GB  of RAM, and a
500 GB disk  (7200 rpm).  The software platform  is composed of Debian
Linux 6.0.5 (kernel 2.6.32)  and Oracle Java 1.6.0\_26 virtual machine
(JVM).  The network switch and  interfaces in each node were dedicated
to the experiments, not  transmitting any other traffic.  Treplica was
configured to use  the local disk of the node  that hosted the replica
as its  persistent data  store, so disk  accesses did not  trigger any
network activity.   These precautions  were adopted to  guarantee that
the  network was  used only  to carry  the messages  exchanged  by the
replicas due to the activity generated by Treplica (Paxos).

\subsubsection{Workload}

For all experiments we run a system with 5 replicas under a fixed load
of  1000 op/s  during 390  seconds.   This workload  is guaranteed  to
remain fixed at  1000 op/s in the presence  of crashes and recoveries,
by  processing  requests  from  clients  only at  the  always  correct
processes.  The first 90 seconds and final 60 seconds are discarded as
ramp up and ramp down time, for a total of 240 seconds of steady-state
run  time.  All values  of time  that appear  in the  text and  in the
figures are  relative to the  beginning of the  steady-state execution
period.  For each of the faultloads described in the next section, and
for each type of validation  procedure (original and seamless) we have
performed  20 distinct runs  and recorded  the average  performance in
operations per second, continuously, for the duration of the run.

\subsubsection{Faultload}

We have  created two faultloads,  each one designed to  guarantee that
Treplica exhibits the  intended behavior in relation to  the election of
coordinators:

\begin{description}
\item[\textbf{Non-stable   failure  detector   faultload:}]   In  this
  faultload,  the  failure detector  selects  as  the coordinator  the
  correct process with the smallest process identifier irrespective of
  its failure history,  that is, the active process  with the smallest
  identifier is  elected coordinator even if this  process crashes and
  recovers much  often than another  process with a  larger identifier
  that has  crashed and  recovered less frequently.   Processes retain
  their identifiers across  crashes. To inject this fault,  we let the
  non-stable failure  detector choose the coordinator;  it chooses the
  process  with the  smallest  identifier.  Next,  the coordinator  is
  crashed  and  restarted. The  non-stable  failure detector  promptly
  re-elects the recently restarted process as the coordinator.

\item[\textbf{Stable failure detector  faultload:}] In this faultload,
  the  failure detector used  is stable,  that is,  it selects  as the
  coordinator the  correct process that is crash-free  for the longest
  time.  This  time, to inject  the fault, the connection  between the
  coordinator and the rest of  the replicas is interrupted. The result
  of  the fault  injection  is a  system  partitioned in  two sets  of
  processes:  (i)  $U$,  a  unitary  set that  contains  the  isolated
  coordinator  and (ii) $R$,  the set  that contains  the rest  of the
  Treplica processes.  As soon  as the partitioning occurs, the stable
  failure detector selects as the new coordinator the stablest process
  among the processes of $R$.  Later, the sets $U$ and $R$ are united,
  by removing the network partition.  After reunification, the process
  coming  from $U$, the  original coordinator,  is still  the stablest
  process among  all the processes.   So, the stable  failure detector
  demotes the  current leader and  selects as coordinator  the process
  that belonged to $U$.

\end{description}

We  have adapted  the coordinator  selection procedure  implemented by
Treplica to allow the deployment of the faultloads described above and
created the following experiment setups:
 
\begin{description}
\item[\textbf{Non-stable failure detector  setup (NFD):}] A non-stable
  failure  detector  is used.   The  JVM  that  hosts the  coordinator
  process is  brought down at  $t$ =  60 s and  remains down for  30 s
  until $t$ = 90 s.  After this period, the crashed process is brought
  back into operation.

\item[\textbf{Stable failure detector setup (SFD):} ] A stable failure
  detector is used.   The network interface of the  computer where the
  coordinator process is  executing is brought down at $t$  = 60 s and
  after 30 s it is brought back up, at $t$ = 90 s.
\end{description}

\begin{figure*}[htb]
  \centering
  \includegraphics[width=13.9cm]{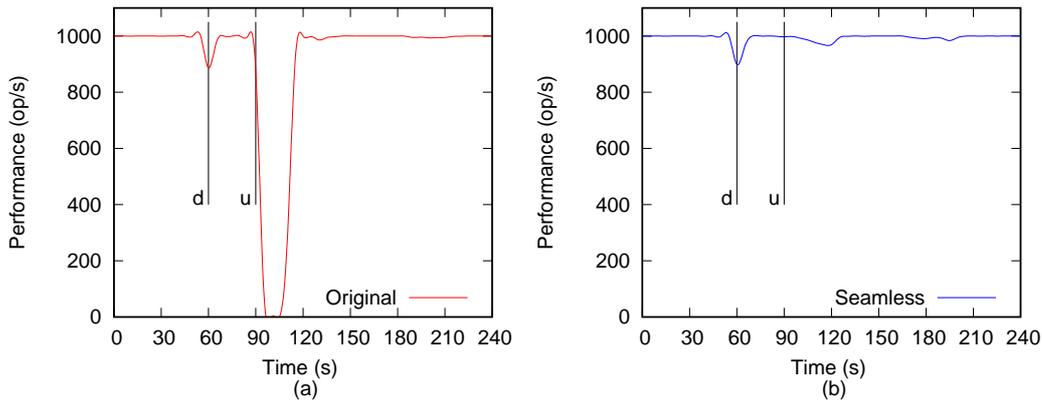}
  \caption{Non-stable Failure Detector Setup (NFD)}
  \label{fig:crash}
\end{figure*}

\begin{figure*}[htb]
  \centering
  \includegraphics[width=13.9cm]{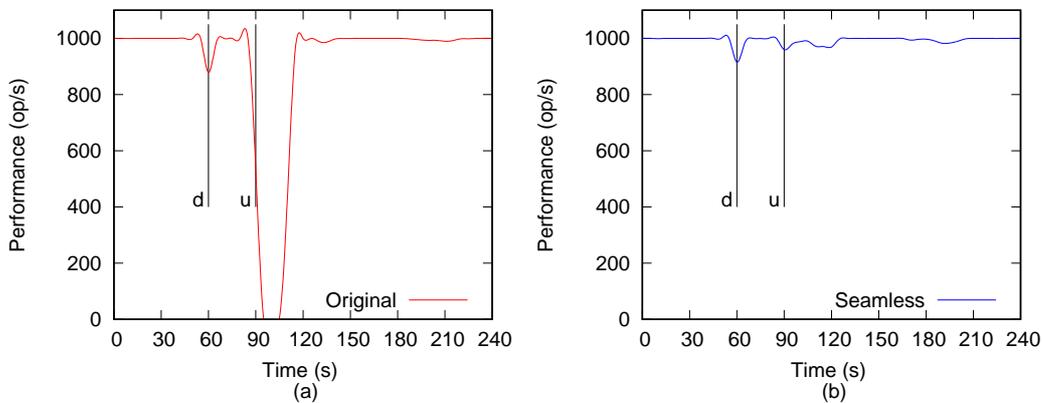}
  \caption{Stable Failure Detector Setup (SFD)}
  \label{fig:net}
\end{figure*}

All faults are injected at the operating system level, using automated
scripts that do not require any human intervention during the duration
of the experiment.   Both faultloads are based on  the assumption that
the failed process is reinstated as  coordinator as soon as it returns
into operation.  As such,  these faultloads emphasize problems arising
from  failure detectors  with inadequate  accuracy/completeness, wrong
implementations or not correctly tuned  parameters.  The recovery of a
replica in the original coordinator validation is performed as fast as
possible.   This  is  done  to ensure  that  the  application  remains
unavailable  for  the  shortest  period possible.   For  the  seamless
coordinator  validation, recovery  is performed  at a  slower rate  to
minimize   its  impact   on  the   performance  of   the  application.
Nonetheless,  in both  cases the  full recovery  of the  coordinator's
state occurs within the duration of the experiment.

\subsection{Results}

\begin{table*}[htb]
  \caption{Average Performance  of the Application  for each Faultload
    Setup}
  \label{table:failure}
    \centering
    \begin{tabular}{lllll} \hline
      Faultload &  Original (op/s) &  COV & Seamless  (op/s) &  COV \\ \hline
      Process failure (NFD) & 917.50 & 0.0078 & 993.40 & 0.0015 \\
      Network partition (SFD) & 912.39 & 0.0091 & 992.97 & 0.0021 \\ \hline
    \end{tabular}
\end{table*}

Figures~\ref{fig:crash}  and  \ref{fig:net} show  the  results of  the
experiments executed  with the NFD  and SFD setups,  respectively.  In
the figures,  vertical lines labeled \emph{d} indicate  the moment the
coordinator process fails or  the network is partitioned. The vertical
lines  labeled  \emph{u} indicate  the  time  the coordinator  process
recovers or the network is restored.

For both  faultload setups we  observe the same general  behavior.  As
expected, the performance  of the application is affected  at both the
moment a replica crashes and the moment it starts its recovery, as can
be seen by the performance drops  around 60 s and then beginning at 90
s.  While  the original coordinator, say process  $p_c$, remains down,
the failure detector elects another coordinator, say $p_c'$, among the
remaining  processes   and  the   execution  of  the   application  is
practically unaffected, as indicated by the relatively brief and small
performance drop observable at t = 60 s.

During  recovery,  a  more  interesting  behavior  emerges,  with  the
throughput  of the  application  based upon  the original  coordinator
validation momentarily  dropping to  0 op/s for  both the NFD  and SFD
setups.   In the NFD  setup, the  recovery of  process $p_c$,  that by
construction has the smallest identifier, induces the failure detector
to  demote the  current coordinator  $p_c'$  in favor  of $p_c$.   The
application then waits for the  recovery of $p_c$ to end, meaning that
from  t  =  90  s  to  t  = 110  s,  the  application  is  unavailable
(Figure~\ref{fig:crash}  (a)).   This is  one  of  the  reasons why  a
non-stable  failure  detector  should  not  be  used  in  practice  to
implement  Paxos  and  some  form  of leader  stability  is  required.
However, stability isn't simple to  define and to achieve.  In the SFD
setup, the coordinator $p_c$ is  isolated from the other processes and
unable  to keep  up with  the remaining  replicas for  30 s.  When the
connectivity is restored, once  again the failure detector demotes the
current coordinator  $p_c'$ because $p_c$ has indeed  been the process
with  the longest uptime  among the  processes, again  the application
halts while  the recovery of $p_c$ is  on course (Figure~\ref{fig:net}
(a)).

Meanwhile, for  both the  NFD and  SFD setups,  the throughput  of the
application implemented  atop of a Treplica  with seamless coordinator
validation was  only slightly reduced,  regardless of the  behavior of
the  failure  detection  mechanism  (Figures~\ref{fig:crash}  (b)  and
\ref{fig:net}  (b)).  In  fact,  the  impact of  the  recovery of  the
coordinator on  the performance of  the application is similar  to the
impact  of  the  coordinator's  failure  on  the  performance  of  the
application,  and  both are  small  and  brief.   The results  of  the
experiments for  the NFD and  SFD setups show  that in both  cases the
Treplica  with  seamless  coordinator  validation  has  prevented  the
application from  becoming unavailable  by maintaining  the throughput
during the failure-affected  periods practically at the  same level of
the throughput measured during the failure-free periods.

Using samples of  the average throughput, we have  determined that the
minimum number of  executions required per experiment  to guarantee an
accuracy of  5\% for  the performance  measurements with  a confidence
level of  99\% is  4~\cite{jain91}. The average  throughput of  the 20
runs  actually   carried  out  for   each  experiment  is   listed  in
Table~\ref{table:failure},  with  the  corresponding  coefficients  of
variation (COV).   The very small COVs  are an extra evidence  that the
average  performance   gain  obtained  by  the   seamless  coordinator
validation  in  comparison  with   the  performance  of  the  original
coordinator   validation,   for   the   same   faultload   setup,   is
significative.

These  results allow  us  to conclude  that  the seamless  coordinator
validation introduced here definitely improves the availability of the
application  supported  by Paxos  in  the  presence of  a  coordinator
failure and recover,  even if the failure detection  mechanism is very
unreliable.

\section{Related Work}
\label{sec:related}

The importance  of the coordinator replacement  procedure was observed
by  Chandra et  al.  during  the design  and operation  of  the Chubby
distributed lock system~\cite{burrows06,chandra07}. In this system the
current  coordinator   has  an  explicit   lease  to  operate   for  a
predetermined period of time. This  coordinator is called a master and
it  uses its  lease to  ensure  its stability  and concentrate  client
requests.  The  designers of  Chubby decided to  make it harder  for a
replica to loose its master status to simplify the design and increase
its  reliability.  However,  this  approach  has the  cost  of  slower
detection of process failures.   For instance, a typical master change
takes around  14 seconds~\cite{burrows06}.

In general,  a way to mitigate  this problem is to  devise a mechanism
that  makes it  harder to  replace  the coordinator,  namely a  leader
stabilization mechanism.  Malkhi  et al.~\cite{malkhi05} have proposed
a failure detector based on an election procedure with built-in leader
stability;  the coordinator  is  only  replaced if  it  isn't able  to
effectively perform its actions.  However, approaches like this do not
directly  address the  problem of  coordinator replacements  caused by
message loss  or variable communication delay.   Minimization of these
errors entails  the improvement in  the overall quality of  service of
the  failure detector~\cite{chen02}, which  often requires  tuning the
detector's parameters  to the characteristics of  the local networking
environment.   In the  absence of  a self  adjusting mechanism  and in
rapidly changing network  conditions, the system has to  bear the full
cost of coordinator replacement more often than necessary.

Ultimately, the  fact that Paxos  requires a single coordinator  is at
the  root of  the unavailability  problem.  This  single  process will
eventually fail,  or be mistakenly  taken for failed, requiring  a new
coordinator to take its place.  Another approach was taken by Camargos
et al.  and consists in not relying  in a single one but on a group of
coordinators~\cite{camargos08}.  Their  justification is that multiple
coordinators make the algorithm more resilient to coordinator failures
without requiring  the use of Fast  Paxos and its  larger quorums. The
resulting algorithm  is considerably complex and  increases the number
of  messages  exchanged  between   the  acceptors  and  the  group  of
coordinators.   Our  seamless   coordinator  validation  procedure  is
simpler  and has similar  coordinator resilience,  if we  consider the
whole set of  replicas that can act as a coordinator  as a group where
only a master is active at any time and master changes are very cheap.

A  similar  strategy of  splitting  the  coordinator  role among  many
processes was taken in Mencius~\cite{mao08}, to minimize the number of
exchanged messages  in a wide-area network. In  Mencius processes take
turns running a coordinator  and proposers only exchange messages with
the closest coordinator.  The handover of coordinator responsibilities
to   another  process   is   a  built-in   feature   of  the   Mencius
protocol. Every  instance has a predefined  coordinator, that proposes
and  decides  a  value  in  it  or decides  a  special  value  (no-op)
indicating  it yields  its turn.   In Mencius  coordinator replacement
occurs on a per-instance basis.  This effectively solves the temporary
unavailability problem, as  the state to be transferred  is reduced to
one  instance.  However,  in  the case  of  a permanent  failure of  a
process, an unbounded number  of these simple coordinator replacements
will happen  continuously for  as long as  the failed  process remains
down.

One  of   the  causes  of  communication   instabilities  that  induce
coordinator replacements in the absence of process failures is message
loss   due    to   buffer   overflows.    The    designers   of   Ring
Paxos~\cite{marandi10} have  observed that many  concurrent senders of
multicast  messages  can increase  considerably  the  rate of  message
loss. Ring  Paxos attacks the  problem caused by these  message losses
from a throughput perspective, by organizing acceptors in a ring. This
minimizes concurrent senders, decreases message loss and increases the
utilization  of  the  links.    However,  in  Ring  Paxos  coordinator
replacement is still  an expensive operation that can  be triggered by
workload peaks; it includes reforming the ring topology and broadcasting
it to all active agents.

\section{Conclusion}
\label{sec:conclusion}

In  this paper  we  have shown  a  novel way  to  avoid the  temporary
unavailability problem  caused by Paxos  coordinator replacements. Our
solution  is based on  the observation  that the  validation of  a new
coordinator is  composed of  two activities: activation  and recovery.
We have shown  that only the completion of  the activation is strictly
required before  the coordinator can resume its  operation.  This fact
has led us to a seamless coordinator validation that has two important
characteristics.   First,  it allows  activation  and  recovery to  be
performed concurrently.   Second, it reduces  the information required
to activate the new coordinator  to a single integer exchanged between
the acceptors.

We  have   verified  experimentally  that   the  seamless  coordinator
validation avoids the temporary unavailability problem in the presence
of  process   crashes,  providing  uninterrupted   operation  for  the
application  built  atop  Paxos.   These  results  indicate  that  our
seamless coordinator validation ensures  that the performance of Paxos
becomes orthogonal  to the reliability  of the failure  detector, that
is, there  is less  need for stable  coordinators. This is  of special
importance to the autonomous  operation of replicated applications, as
even  the more  finely tuned  failure detection  mechanism  can, under
changing   network  conditions   and  failure   patterns,   behave  in
undesirable ways.  As such,  a validation mechanism that is impervious
to the  error rate of the unreliable  failure detection implementation
is highly desirable.

\begin{acknowledgements}
  Gustavo  M.   D.   Vieira  was partially  supported  by  CNPq  grant
  142638/2005-6.   Luiz E.   Buzato  was partially  supported by  CNPq
  grant 473340/2009-7 and FAPESP grant 2009/06859-8.

  The authors  thank Prof. W.  Zwaenepoel, and Olivier  Cramieri, both
  from EPFL, Switzerland, for their  support in the earlier stages of
  this  research. We  thank  Daniel  Cason for  the  support with  the
  cluster management at IC-UNICAMP.
\end{acknowledgements}

\bibliographystyle{spmpsci}
\bibliography{paxos-coordinator}

\end{document}